# Circular Dichroism of Single Particles

Eitam Vinegrad[1], Daniel Vestler[2], Assaf Ben-Moshe[2], A. Ronny Barnea[2], Gil Markovich[2] and Ori Cheshnovsky[2,*]

1. School of Physics, Raymond and Beverly Faculty of Exact Sciences, Tel Aviv University, 69978 Tel Aviv, Israel.
2. School of Chemistry, Raymond and Beverly Faculty of Exact Sciences, Tel Aviv University, 69978 Tel Aviv, Israel.

*Corresponding author: orich@post.tau.ac.il

August 16, 2017

**Circular dichroism (CD), induced by chirality, is an important tool for manipulating light[1,2] or for characterizing morphology[3] of molecules[4], proteins[5], crystals[6] and nano-structures[7–11]. CD is manifested over a wide size-range, from molecules to crystals or large nanostructures. Being a weak phenomenon (small fraction of absorption), CD is routinely measured on macroscopic amount of matter in solution, crystals, or arrays of fabricated meta-particles. These measurements mask the sensitivity of CD to small structural variation in nano-objects. Recently, several groups reported on chiroptical effects in individual nanoscale objects: Some, using near-field microscopy, where the tip-object interaction requires consideration[12,13]. Some, using dark field scattering on large objects[14,15], and others by monitoring the fluorescence of individual chiral molecules[16,17]. Here, we report on the direct observation of CD in individual nano-objects by far field extinction microscopy. CD measurements of both chiral shaped nanostructures (Gammadions) and nanocrystals (HgS) with chiral lattice structure are reported.**

CD measurements of individual particles or structures require a supporting substrate. This constraint must be carefully considered while attempting to measure and analyze the resulting signal. The presence of a substrate may transform an otherwise achiral structure into a truly chiral entity. Even more significant is the phenomenon of extrinsic chirality that can emerge even for truly achiral objects, such as effects due to edges or when the excitation beam is not perpendicular to the plane of examined object. All these may induce chirality or even linear optical activity disguised as circular chiroptical effects. We will show that the axis of the excitation beam should be perpendicular to the substrate within 20 μrad to prevent appearance of strong tilting-induced extrinsic chirality.

Our experimental setup for spectroscopic measurements of individual nanostructures follows the approach of Sandoghdar and coworkers for measuring the absorption of single molecules[18] and nanocrystals[19]. Our system, depicted in Figure 1, is equipped with a broadband light source (Fianium Supercontinuum source, SC). The excitation beam, with wavelengths in the range of 450-720 nm is selected using an acousto-optic tunable filter (Gooch & Housego, AOTF) and passed through accurate polarization control (liquid crystal retarder, LC) to enable CD spectroscopy. The circularly polarized light is focused on the sample plane as we scan through both the spectral range as well as over the sample of interest and preform spectroscopic extinction and reflectance measurement on a single particle. The LC retarder cycles 10-50 times between Left and Right circular polarization states to measure the difference between the corresponding absorptions, $A_{LCP} - A_{RCP}$, where $A_{LCP}$ and $A_{RCP}$ are the measured extinctions of the left and right circular polarizations, respectively. The dissymmetry is defined as $g = \frac{A_{LCP} - A_{RCP}}{A_{avg}^{Max}}$, where $A_{avg}^{Max}$ is the average extinction at the center of the particle.

In order to compensate for source intensity variations we use a 50:50 beamsplitter to split the light to a probe beam and a reference beam. The reference beam is used to monitor light intensity variations while the probe beam is passed through the LC retarder and then focused using a 0.85NA polarization maintaining objective. A key component for an accurate CD measurement is the 2 axis tilt sample holder, which enables the adjustment of the sample to be perpendicular to the beam axis within 20 µrad.

Dissymmetry spatial measurements were performed by raster scanning an area containing a single nano-object, with 100-150 nm steps using a fixed wavelength and cycling between polarization states. Dissymmetry spectra in the range of 470-720 nm were measured at the geometric center of the nano-object after determining the center position from a transmission scan minimum. To account for residual asymmetry of the measurement system ($<10^{-4}$) in respect to the two circular polarizations, our measurements were normalized to an equivalent measurement taken on the bare substrate a few microns away from the particle.

Our first set of single nano-object CD measurements was performed on fabricated chiral shapes for which optical simulations are readily made. Spatial dissymmetry maps at a wavelength of 600 nm of right and left handed, 150 nm wide Gammadions, as well as of a 1000 nm wide square are shown in Figure 2. The dissymmetry shown in Figure 2 are of the order of $g \sim 10^{-3}$. We compare our findings to FDTD simulations (Lumerical Inc) using the experimental geometry determined from SEM images. For the FDTD simulation we use a Gaussian linearly polarized focused light source, running two simulations, one for each orthogonal linear polarization. The right and left handed circular polarization response was generated by summation of the electric and magnetic fields of the two simulations, using the appropriate phases[20]. The simulated Gaussian beams were focused just few a nanometers in front of the Gammadion glass/ITO[21] interface, as in the experiment. For the dielectric function of gold we used the published data by Rakić et al[22] .

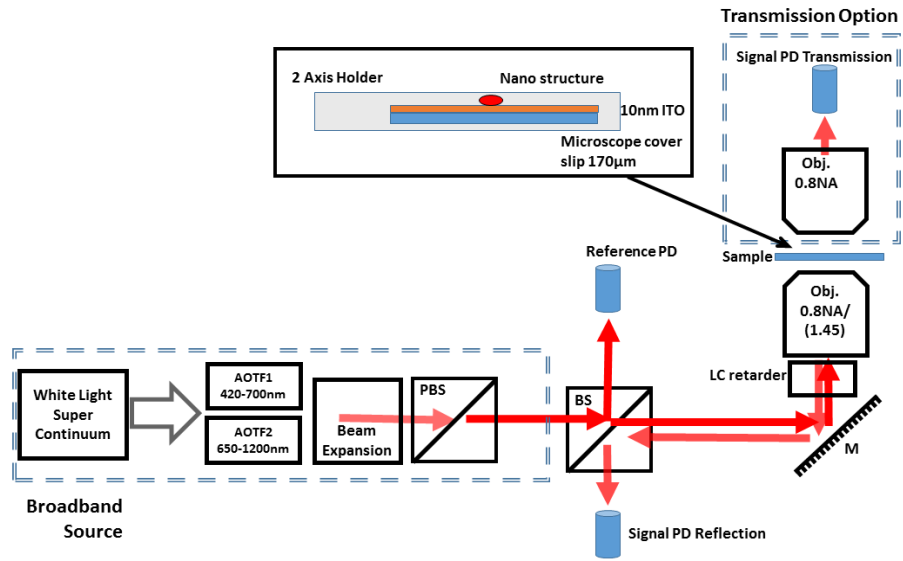

*Figure1 – Schematic of the experimental setup. A broadband supercontinuum source, filtered by an AOTF, is passed through a 50:50 beamsplitter separating the light to a reference beam and a probe beam. The probe beam polarization is controlled by the LC retarder and then focused onto the sample. The transmitted and reflected light are both collected and intensity balanced, using the reference beam.*

Note in Figure 2 the dissymmetry sign change between the left and right handed Gammadion configurations at 600 nm wavelength. In our diffraction-limited experiment, the transmission dissymmetry signal coincides with the geometric center of the particle. This is in agreement with our simulations for the spatial distribution of the dissymmetry signal shown in Figure 3d.

For the achiral square shape, the dissymmetry spatial distribution is more complex. It is characterized by four lobes surrounding the square, with dissymmetry values comparable to those of the Gammadions. These values are diminishing, as expected, towards zero at the shape's center. The fact that two of these lobes are positive and two are negative effectively cancels the integral of the CD across the shape, so the total dissymmetry from the achiral shape is close to zero.

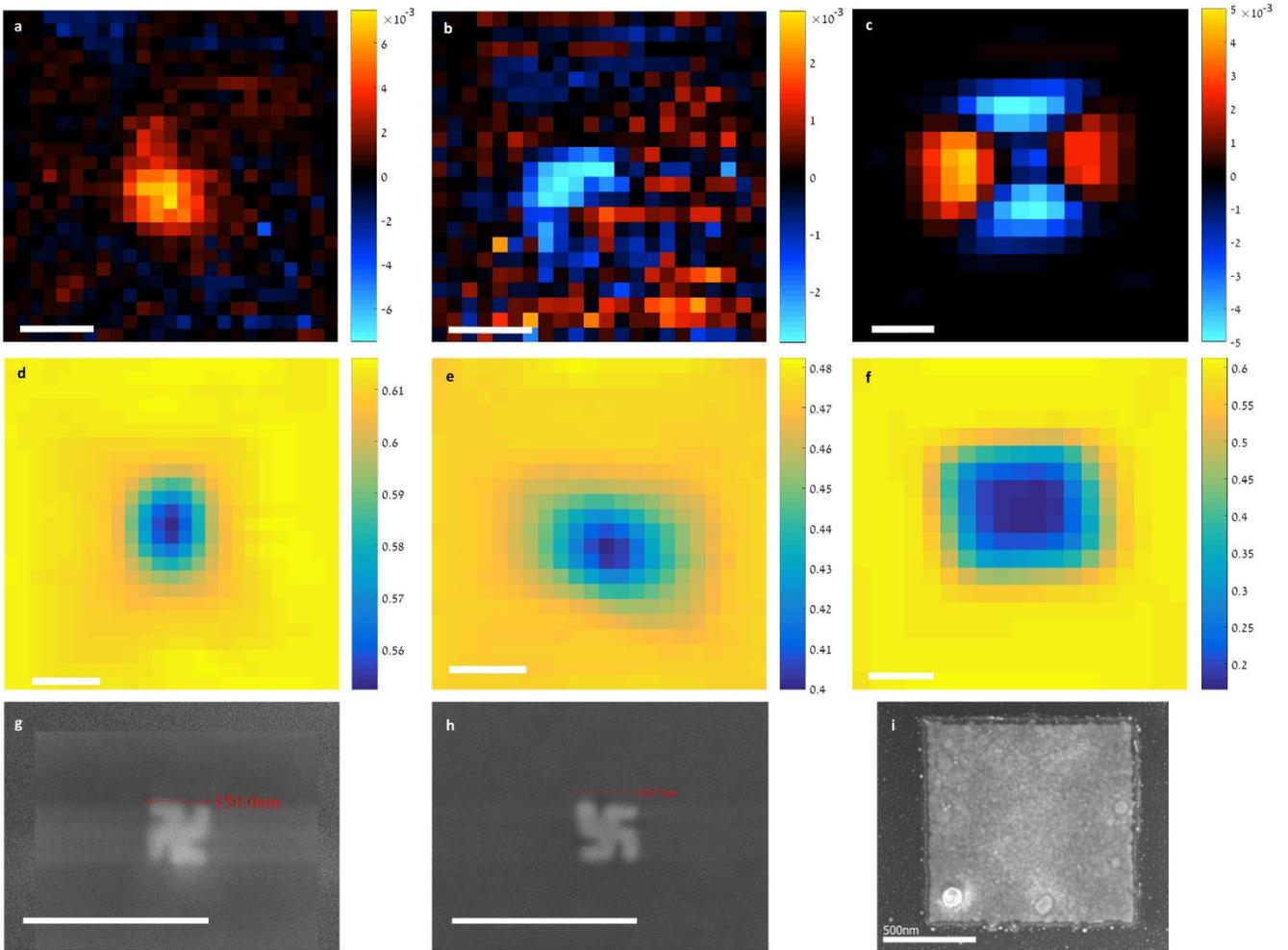

*Figure 2 - Spatial dissymmetry maps taken in transmission mode using a 600 nm wavelength. (**a**) 150 nm left handed Gammadion. (**b**) 150 nm right handed Gammadion. (**c**) 1 µm square. (**d-f**) the corresponding transmission maps respectively. (**g-i**) the corresponding SEM micrographs. The white bars represent 500 nm. The sample were scanned using 100 nm steps.*

Intrigued by the apparent dissymmetry signal for the edges of an achiral shape, we studied this phenomenon by using a series of FDTD simulations, in which we scan a 1 µm square along one of its symmetry axes across the Gaussian beam in horizontal or vertical directions. We also investigated the influence of sample tilt with respect to the focal plane. The simulation results are depicted in Figure 3a-c. Note that the dissymmetry sign changes between the horizontal and vertical directions in agreement to our measurement. We learnt that even a slight sample tilt of 20 mrad can enhance the dissymmetry[23,24] by 3 orders of magnitude. This effect should be carefully considered in all CD measurements, as any type of symmetry breaking increases the measured dissymmetry, be it un-uniformity in the light beam or a slightly slanted transition from the elevated shape to its surroundings. In experiment, as in our FDTD simulations on achiral square, all contributions to the shape dissymmetry arise from the interface between the shape and its surroundings and converge to zero at the center.

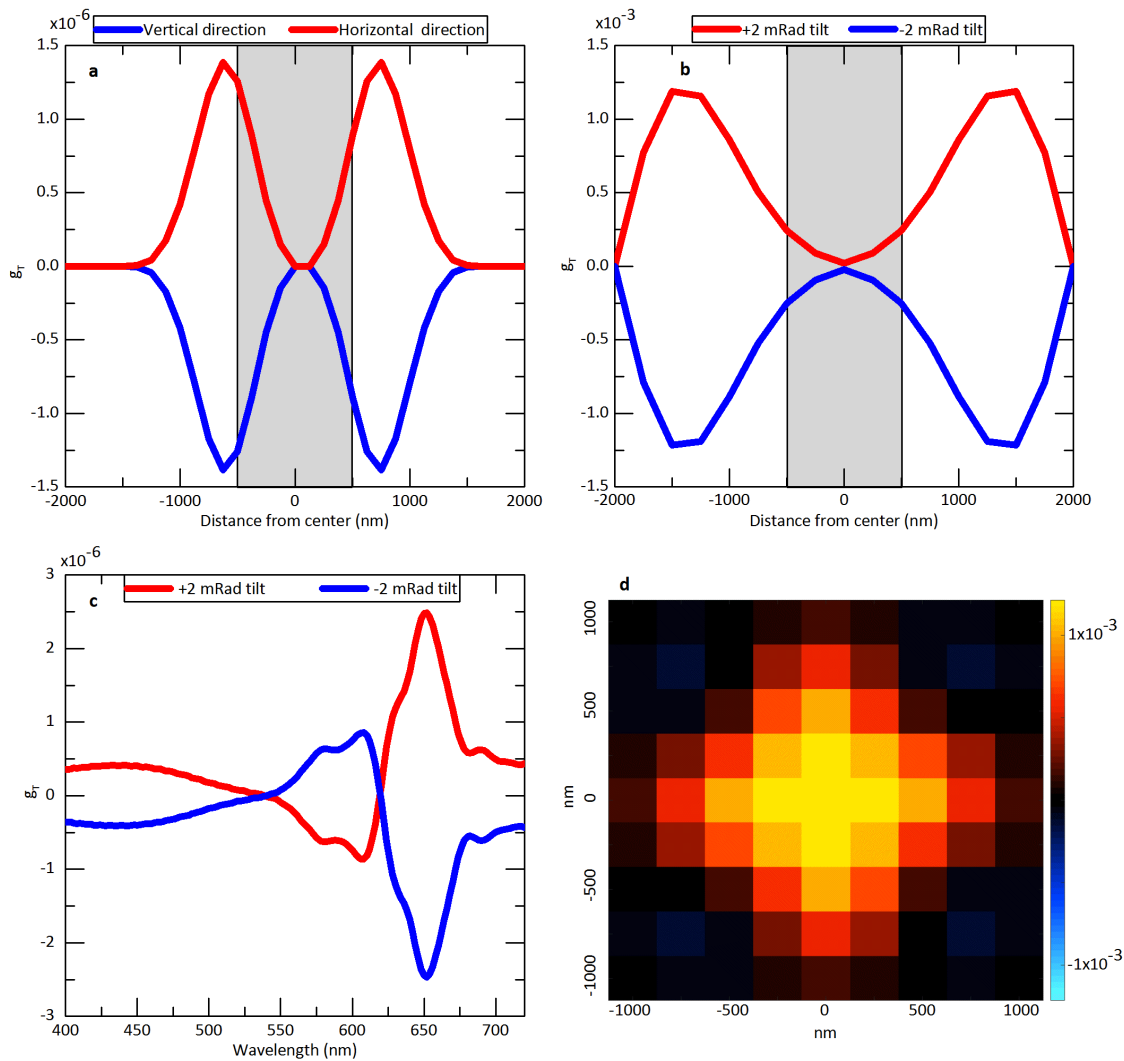

*Figure 3 – FDTD simulations, dissymmetry spatial distribution and tilt dependence for an achiral 1μm square. (a) Scan in the horizontal and vertical directions. (b) Scan in the horizontal direction with a ±2 mrad tilt. (c) Dissymmetry spectra at the point of maximum dissymmetry for a ±2 mrad tilt. (d) Transmission dissymmetry spatial distribution FDTD simulation for a 500 nm Gammadion at a wavelength of 650 nm.*

Spectroscopy measurements were performed at the center of Gammadions with total width of 550 nm and a line width of 75 nm. With careful adjustment of the excitation beam to be perpendicular to the sample plane within 20 μrad, the CD signal is spatially centered at the Gammadion geometric center (see supplementary Figure 1 for more information on the adverse effects of misalignment). The center was located by finding the point with lowest transmission (using 20 nm steps). At this location, the polarization was modulated back and forth between right and left circular polarizations for 30 cycles. Figure 4 depicts the measured spectra for both right and left handed Gammadions as well as the corresponding FDTD simulated spectra.

We note that the measured spectra of the right and left handed Gammadions indeed show opposite CD signs as we expected for both transmission as well as for reflection measurements (see supplementary Figure 5). The experimental spectrum is in good agreement with simulations. The two enantiomers show slightly different dissymmetry line shapes, as well as a higher variance in the spectra for the three different right handed Gammadions. This can be attributed to variance of the structures created using the e-beam lithography. We also note the spatial dissymmetry maps show the opposite signal distributed around the Gammadion geometric center, in agreement with the simulated spatial distribution depicted in Figure 3. The maximum value for each of these spatial dissymmetry maps matches the values of the measured dissymmetry spectra at the coresponding wavelength.

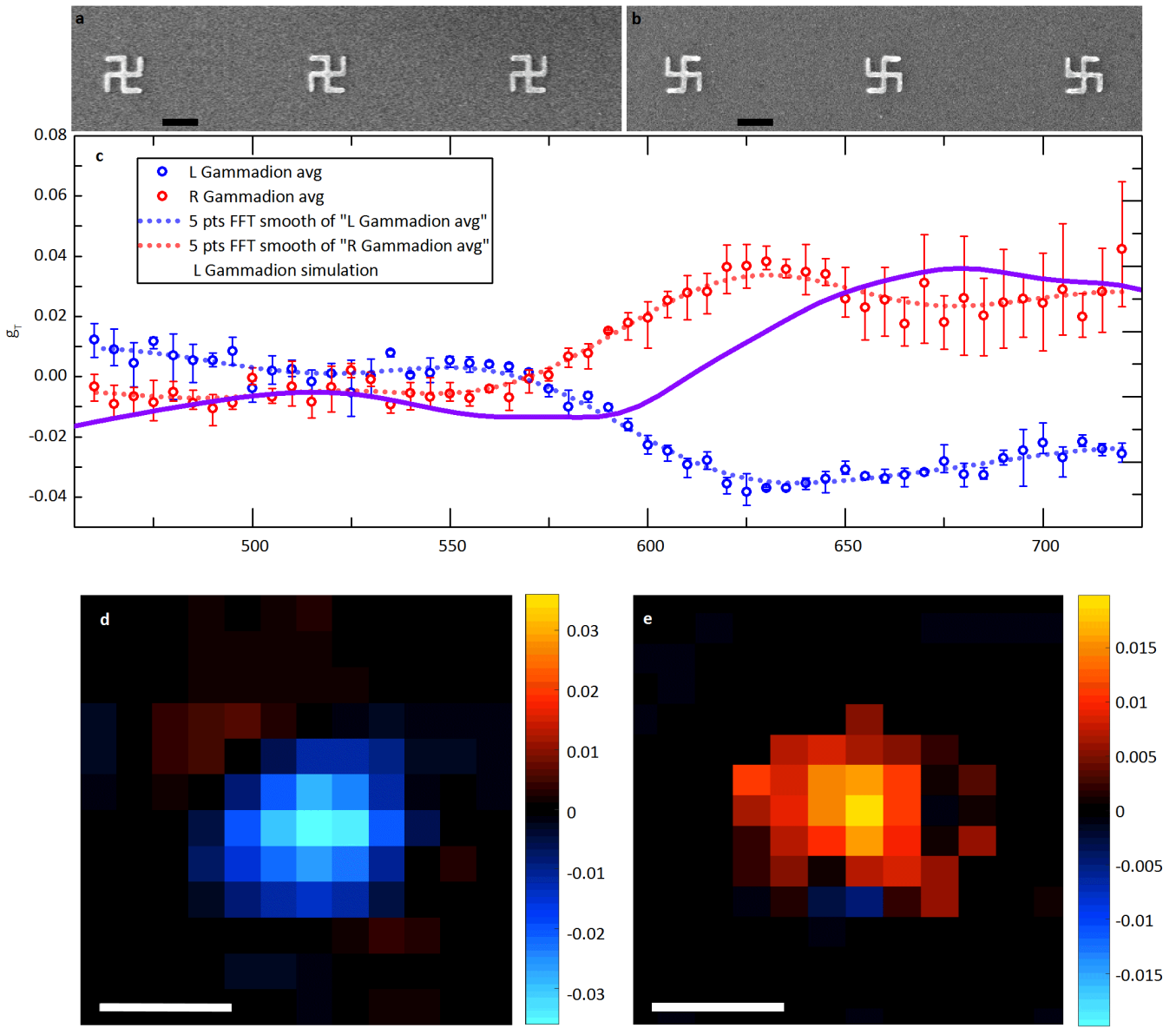

*Figure 4 – (a, b) –SEM micrographs of the left and right handed 550 nm wide gammadions for which we performed spectroscopic measurements (scale bar 1 µm). (c)- Transmission dissymmetry spectra for 550 nm Gammadions. The bars represent the variation over the three different samples, the hollow circles represent the average of the three measurements, the dotted line is the average filtered by a low pass FFT filter while the solid purple line is the simulated spectra calculated using FDTD for a Gaussian beam focused a few nm bellow the interface at the geometrical center of the shape. (d, e) Left handed (at 650 nm wavelength) and right handed (at 600 nm wavelength) 550nm Gammadions spatial dissymmetry maps, scale bar represents 500 nm.*

After studying chirality expressed in a geometrically chiral meta-particle, we examined chirality originating from a chiral crystal lattice arrangement. The system of choice was colloidal chiral HgS nanoparticles, grown to an average size of 100 nm. As evident in the TEM micrograph, Figure 5, the geometry of the nanocrystals is achiral, which greatly simplifies the resulting CD signal interpretation. Samples were prepared by colloidal growth of HgS nanocrystals in solution with the presence of D-penicillamine[25,26]. A dilute solution of particles was drop cast on a microscope cover glass. We have identified individual particles and measured their dissymmetry at the particles geometric center as determined by the maximum of the measured extinction. Since the extinction of HgS nanoparticle is quite low, extraction of exact extinction values from two sequential measurements (for background and over a particle) often led extinction values to oscillate around zero, this led to non-physical dissymmetry values (i.e. g>1). In order to describe the optical activity in a meaningful way, we therefore used the normalized difference in transmission intensity $g^* = \frac{I_{RCP}-I_{LCP}}{I_{RCP}+I_{LCP}}$, for low extinction values. This preserves the dissymmetry spectral line shape and only changes the scale. Figure 5 shows $g^*$ measured for both aggregates and a single HgS particle in comparison to ensemble measurements done on the stock solution of these particles using a commercial CD spectrometer. Both spectra, of aggregates and the single particle, are in good agreement with the ensemble spectra.

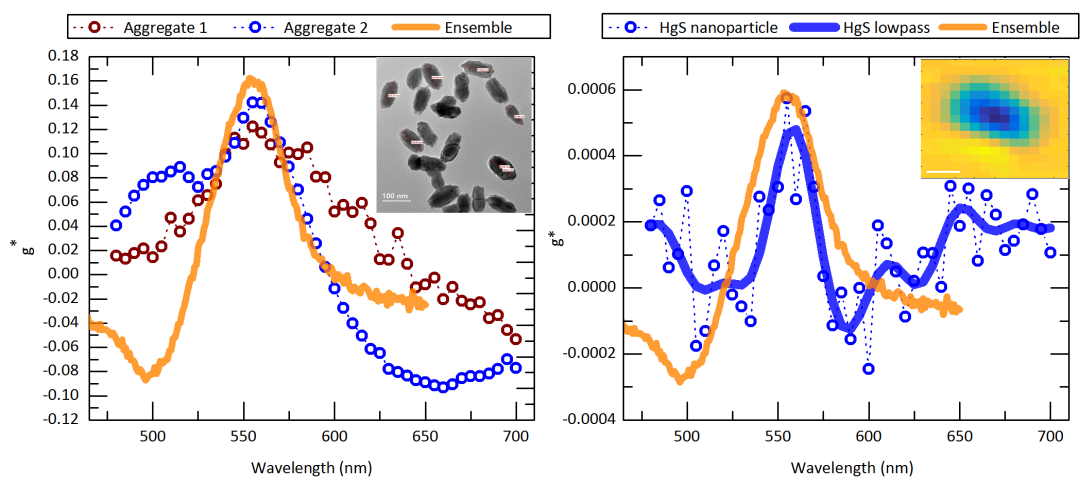

*Figure 5 - Left- $g^*$ spectra of HgS aggregates vs the solution measurement. Inset - TEM micrograph showing the typical nanoparticle size and geometry. Dots represent the microscope dissymmetry measurements on two different aggregates, the solid orange line represents the dissymmetry measured on the stock solution of these particles. Right- dissymmetry spectra of a single HgS nanoparticle. Inset - Transmission scan for the same particle. Hollow circles represent dissymmetry on a single HgS nanoparticle, blue solid line represent the results from the single particle filtered using a low pass and solid orange line represent the dissymmetry measured on the stock solution of these particles. The ensemble dissymmetry was rescaled by a constant factor to bring it to have the same maxima as the $g^*$ data.*

We have demonstrated CD measurements of single nano-objects, both for geometrically induced chirality as well as for chirality originating from a chiral crystal space group. We have shown that the spectral features of the single particle are comparable to those of the ensemble, as expected. We have also shown the excessive sensitivity of CD measurements to optical alignment and polarization control. On axis illumination within 50 µrad and perpendicularity within 20 µrad are required when measuring CD of individual nanostructures. We estimate our sensitivity to be of the order of $10^{-5}$ in extinction difference between the right and left circular polarizations. Further sensitivity as well as the ability to measure smaller particles, or particles with lower dissymmetry values, could possibly be achieved by using a higher frequency modulation scheme, where noise will be lower, e.g. using an electro optic modulator to switch between two linear polarizations and a fixed quarter wave-plate retarder just before the focusing objective.

## Methods

**Materials.** Mercury (II) Nitrate monohydrate and Thioacetamide 98% were purchased from Sigma-Aldrich, D-penicillamine was purchased from Fluka and Sodium Hydroxide pellets were purchased from Frutarom. All materials were used as is to produce the relevant solutions.

**Sample preparation.** The 550 nm wide Gammadions were fabricated using e-beam lithography on top of a 1.5H microscope cover-slide coated with a 10 nm adhesion layer of ITO. The Gammadions were ~27 nm in height (2 nm Chromium and 25 Gold) with a line width of 75 nm. The 150 nm wide Gammadions were fabricated using focused ion beam milling of a 15 nm thick Gold layer on top of a 2 nm thick Chromium adhesion layer on a 1.5H cover-slide. The ion beam milling left the Gammadion shape in the middle of a clear 2 µm window.

**Mercury Sulfide nanoparticle synthesis.** HgS seeds sized 5-10 nm were prepared as described by Ben-Moshe et. al[6]. Growth of the seed particles to 100 nm diameter was conducted in a step by step synthesis similar to the method describe by Ouyang et. al[26]. Each step begins with taking 2 ml of the seed solution and cleaning by centrifuge (Starting with 6000 RPM for 5 mins and decreasing RPM as the particles grow in size). The precipitant is dissolved in 5 ml triple distilled water. 600 µl $Hg(NO_3)_2$ 100 mM, 300 µl D-Penicillamine 100 mM and 200 µl NaOH 1M were then added to the seeds solution under vigorous stirring and the vessel was placed on a hot plate set to 65°C and allowed to stir for 15 mins. 1200 µl of Thioacetamide 100mM were then added to the stirring solution dropwise using a kdScientific syringe pump at a rate of 0.6ml/hr. The solution turns black black and stays that way throughout the dripping process and the resulting particles are larger than the precursor seeds. This process is repeated until the desired size of 100 nm is achieved.

**Experimental setup.** The output of a supercontinuum laser (SC400 4W, Fianium) was filtered by an acousto-optic tunable filter (Gooch & Housego). The output was then polarization-cleaned by a broadband polarizing beamsplitter (PBSH, CVI Melles Griot). The beam is then split by a 50:50 non polarizing beamsplitter into a probe and reference beam. The probe beam passed through a liquid crystal retarder (meadowlark optics) which was carefully calibrated[27] across the visible range. The LC retarder was used to alternate between right and left handed circular polarizations. The probe beam was then focused using a 0.85NA polarization objective (Plan NEOFLUAR 0.85 POL, Zeiss) and collected using another 0.8NA polarization objective (TU plan 0.8 POL, Nikon). The transmitted and reflected light were both focused onto a confocal pinhole and collected by photodiodes (Two Color Sandwich Detectors, OSI Optoelectronics) coupled to a low noise homebuilt trans-impedance amplifier. The light collected by the transmission/reflection diodes was normalized with the light intensity of the reference beam. Samples were raster-scanned with 100 nm steps and dwell time of 30 ms. For dissymmetry measurements the polarization was cycled 30 times through right and left handed circular polarizations, at each scan point.

**Numerical simulations**. These were performed using a commercial software "FDTD solutions" by Lumerical Inc. We performed two separate simulations of the response of the system with two orthogonal linear polarized Gaussian sources, one in each simulation. The results were super-positioned with the appropriate phase to generate the response to the left and right circular polarizations[20]. The sources were simulated using a thin lens approximation and were focused 5 nm before the Gammadion/ITO interface in respect to the light propagation direction. The transmitted and reflected light intensities were calculated using far-field projections at the appropriate numerical aperture for each of the circular polarizations and used to calculate the transmission and reflection dissymmetry. Spatial distribution of the dissymmetry was obtained by moving the Gammadion in 150 nm wide steps in a raster-scan manner, totaling 7x7 points for 150 nm Gammadions and 9x9 for the 550 nm wide Gammadions. The simulation region had PML boundaries and a denser mesh was used in the vicinity of interfaces for obtaining accurate results. Auto-shutoff was set to $1.3 \times 10^{-7}$ fraction of the injected field intensity. The Gammadions dimensions were set to 30 nm height (25 nm Au and 5 nm Cr) on top of a 10 nm ITO layer covering the glass (constant index of 1.5) substrate. All external surfaces were covered with a 3 nm water layer.